\documentclass[conference]{IEEEtran}
\IEEEoverridecommandlockouts
\usepackage{cite}
\usepackage{amsmath,amssymb,amsfonts}
\usepackage{algorithmic}
\usepackage{graphicx}
\usepackage{textcomp}
\usepackage{xcolor}

\usepackage{hyperref}
\usepackage{cite}
\usepackage{latexsym}
\usepackage{txfonts}
\usepackage{here}
\usepackage{booktabs}
\usepackage{multirow}
\usepackage[bottom]{footmisc}
\usepackage[linesnumbered, noend,ruled]{algorithm2e}
\usepackage{url}

\usepackage{subcaption}

\hypersetup{
  colorlinks,
  citecolor=blue,
  linkcolor=blue,
  urlcolor=blue}

\newcommand{\methodname}{Griffin}
\newcommand{\evalBaseline}{\texttt{B+-tree}}
\newcommand{\evalProposal}{\texttt{Griffin}}

\newcommand{\methodnameFull}{Griffin}

\newcommand{\Search}{\texttt{Lookup}}

\newcommand{\Scan}{\texttt{Scan}}
\newcommand{\Insert}{\texttt{Insert}}
\newcommand{\Delete}{\texttt{Delete}}
\newcommand{\Commit}{\texttt{Commit}}
\newcommand{\Abort}{\texttt{Abort}}

\newcommand{\Oconst}{$O(1)$}
\newcommand{\OlogN}{$O(\log N)$}

\newcommand{\YCSBATPS}{3.1}
\newcommand{\YCSBAThreads}{64}
\newcommand{\YCSBETPS}{5.4}

\newcommand{\Lu}{$L_u$}
\newcommand{\Lp}{$L_p$}

\newcommand{\ExperimentPDF}[3]{%
  \begin{figure}[!t]
    \centering
    \includegraphics[width=0.35\textwidth]{artifacts/benchmark/#1}
    \caption{#2}
    \label{fig:#3}
  \end{figure}
}

\newcommand{\ExperimentThreePDF}[9]{%
  \begin{figure*}[t]
    \centering
    \begin{minipage}[t]{0.33\hsize}
        \centering
        \includegraphics[keepaspectratio, width=0.98\hsize]{artifacts/benchmark/#1}
        \vspace{4pt}
        \caption{#2}
        \label{fig:#3}
    \end{minipage}
    \begin{minipage}[t]{0.33\hsize}
        \centering
        \includegraphics[keepaspectratio, width=0.98\hsize]{artifacts/benchmark/#4}
        \vspace{4pt}
        \caption{#5}
        \label{fig:#6}
    \end{minipage}
    \begin{minipage}[t]{0.32\hsize}
        \centering
        \includegraphics[keepaspectratio, width=0.98\hsize]{artifacts/benchmark/#7}
        \vspace{4pt}
        \caption{#8}
        \label{fig:#9}
    \end{minipage}
    \vspace{6pt}
  \end{figure*}
}

\begin{document}


\title{Griffin: Fast Transactional Database Index \\ with Hash and B+-Tree}

\author{
    \IEEEauthorblockN{Sho Nakazono$^1$}
    \IEEEauthorblockA{
      \textit{Computer and Data Science Laboratories} \\
      \textit{NTT}  \\
      Tokyo, Japan \\
    }
    \vspace{10pt}
    \IEEEauthorblockN{Hideyuki Kawashima}
    \IEEEauthorblockA{
      \textit{Faculty of Environment and Information Studies} \\
      \textit{Keio University}  \\
      Kanagawa, Japan \\
    }
    \and
    \IEEEauthorblockN{Yutaro Bessho$^2$}
    \IEEEauthorblockA{
      \textit{Computer and Data Science Laboratories} \\
      \textit{NTT}  \\
      Tokyo, Japan \\
    }
    \vspace{10pt}
    \IEEEauthorblockN{Tatsuhiro Nakamori}
    \IEEEauthorblockA{
      \textit{Faculty of Environment and Information Studies} \\
      \textit{Keio University}  \\
      Kanagawa, Japan \\
    }
}

\maketitle

\footnotetext[1]{Currently, LY corporation.}
\footnotetext[2]{Corresponding author: Yutaro Bessho (e-mail: yutaro.bessho@ntt.com).}

\setcounter{footnote}{2}

\begin{abstract}
Index access is one of the dominant performance factors in transactional database systems.
Many systems use a B+-tree or one of its variants to handle point and range operations.
This access pattern has room for performance improvement. 
Firstly, point operations can potentially be processed in \Oconst{} with a hash table.
Secondly, to ensure serializability of transactions, range operations incur overhead from phantom avoidance techniques
that involve additional processing or synchronization, such as an extra traversal of the B+-tree.
To address these issues, we propose a hybrid index architecture, \methodname{}.
For point operations, \methodname{} has a hash table that provides access paths in \Oconst{} time, along with a B+-tree.
For phantom avoidance, \methodname{} employs a \textit{precision locking} method, which does not involve additional traversal of the B+-tree.
Despite its hybrid architecture, \methodname{} transparently provides linearizable operations and an interface of a single database index.
We built a \methodname{} index combining a hash table and BwTree. Compared to a baseline index that is composed of a BwTree only, it achieves up to \YCSBATPS{}x higher throughput in a point operation dominant workload, and up to \YCSBETPS{}x higher throughput in a range operation dominant workload.
\end{abstract}

\begin{IEEEkeywords}
Database Index, Transaction Processing, Phantom Anomaly
\end{IEEEkeywords}

\section{Introduction}
\label{sec:introduction}

\subsection{Motivation}
Data-driven scientific applications require massive amounts of data, and thus, indexing has been studied to accelerate access to data~\cite{fastquery, insitu}. For example, large scientific data is usually managed with distributed file systems on a cluster of machines, and the performance of modern distributed file systems depends on that of its metadata server~\cite{hopsfs, cfs, DBLP:journals/ngc/TatebeHS10}. Since accesses to the server are transactional, accelerating transactional index has been studied. Transactional indexes are required to handle both \textit{point operations} and \textit{range operations} efficiently.

The design of efficient index structures has been studied extensively in database systems literature, and it has been shown that index structures have a crucial effect on performance.
A study shows that index access constitutes 14-94 \% of the total processing time in a transactional database~\cite{DBLP:conf/micro/KocberberGPFLR13}.
The most commonly used index data structure is the B+-tree.
According to an open database systems catalog~\cite{dbdbio}, 60 out of 87 systems employ a B+-tree or one of its variants.
The conventional practice for adopting a B+-tree is
to have it provide \OlogN{} access paths for both point operations and range operations, where $N$ denotes the total number of elements (i.e., table size).
This design pattern has room for performance improvement.

\subsection{Issues of B+-tree}

\textbf{Issue 1: Point Operation Performance.}
As point operations do not inherently require that the keys be ordered inside the data structure, they can potentially be processed in \Oconst{} time with a hash table \cite{DBLP:books/daglib/0023376}.
However, simply replacing the B+-tree with a hash table significantly hurts range operation performance.
The unordered nature of a hash table demands a full scan regardless of the selectivity of the operation.
Moreover, supporting range operations in transactions requires the elimination of phantoms~\cite{DBLP:journals/cacm/EswarranGLT76}, a type of correctness violation in transactions.
To our knowledge, no techniques have been proposed to eliminate it with a hash table alone.

\textbf{Issue 2: Range Operation Performance.}
The techniques widely used to eliminate phantoms can impose significant overhead on range operations, such as scans.
To avoid phantoms, modern systems use a validation-based method called rescanning.
For example, in Hekaton~\cite{DBLP:journals/pvldb/LarsonBDFPZ11} and Silo~\cite{DBLP:conf/sosp/TuZKLM13}, when a transaction tries to commit, it repeats the record scans it has done to validate that the scanning results have not changed.
The transaction is aborted when any change is detected, preventing phantoms.
This method doubles the calculation costs of scans, each of which consists of a vertical (navigating to a first leaf node: \OlogN{}) and horizontal (scanning through the read set: $O(M)$, where $M$ is the read set size) tree traversal.

\subsection{Proposal}
\begin{figure}[tb]
  \centering
    \centering
    \includegraphics[height=5.4cm]{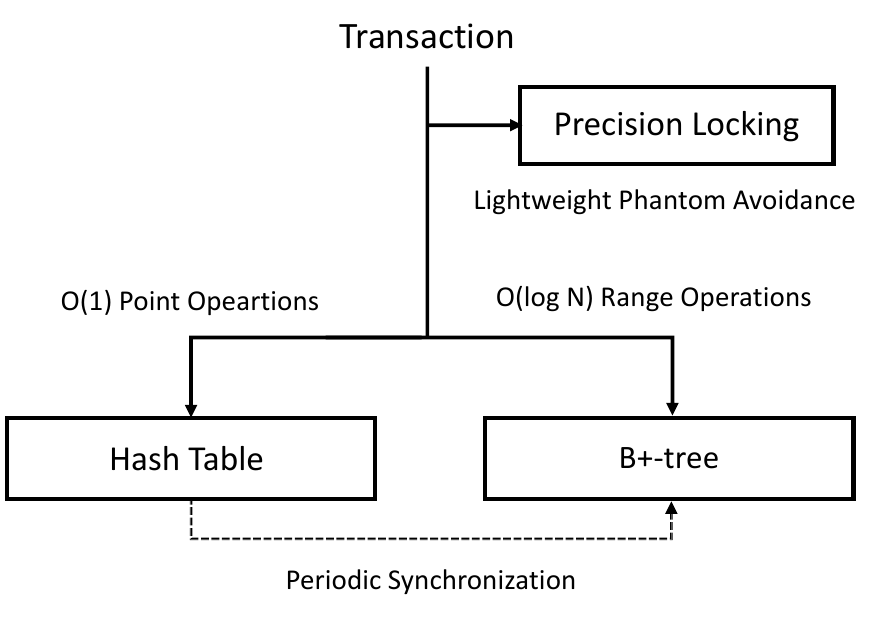}
    \caption{\methodname{} index architecture.}
\label{fig:proposal_architecture}
\end{figure}

To address these performance issues, we propose an indexing architecture called \methodnameFull{}. Fig.~\ref{fig:proposal_architecture} summarizes the architecture of \methodname{}.
The \methodname{} index consists of a hash table and a B+-tree.
As opposed to the singular B+-tree architecture, 
the hash table addresses Issue 1 by providing access paths for point operations in \Oconst{} time.

\methodname{} addresses Issue 2 by avoiding phantoms without rescanning.
\methodname{} adopts a lightweight phantom avoidance technique based on precision locking~\cite{DBLP:conf/sigmod/JordanBB81}.
Inserts and deletes post their access keys to a bookkeeping set shared among transactions.
Before a scan accesses the B+-tree, it searches the set for any conflicting insert or delete keys that conflict with the scan range.
The scan is allowed to proceed if no conflict is found, or aborted otherwise.
This method is advantageous in that it does not involve an extra B+-tree traversal.

\methodname{} provides an interface of a single database index despite the hybrid architecture.
\methodname{} integrates the different index structures and precision locking in a manner that all operations are linearizable~\cite{DBLP:journals/toplas/HerlihyW90}.
Updates (inserts and deletes) take effect on the hash table first, and then periodically get propagated to the B+-tree.
Even though the B+-tree does not always reflect all recent updates, the staleness is mitigated with a synchronization thread that periodically propagates updates to the B+-tree. 
Remaining stale (non-linearizable) reads are detected and aborted, by taking advantage of the precision locking-based phantom avoidance mechanism.

We built an index based on the \methodname{} architecture with a lock-free hash table~\cite{DBLP:journals/dc/GaoGH05} and OpenBwTree~\cite{DBLP:conf/sigmod/WangPLLZKA18}, an open-source implementation of Bw-Tree~\cite{DBLP:conf/icde/LevandoskiLS13a}.
We compared the performance of \methodname{} with a baseline system, which is a B+-tree variant (Bw-Tree) that avoids phantoms through rescanning.
We found that, in a workload that consists of point operations only (YCSB-A), \methodname{} resulted in a peak throughput \YCSBATPS{}x as high as the baseline. We also observed a peak throughput \YCSBETPS{}x as high as the baseline, in a mixture of point and range operations (YCSB-E).

\subsection{Organization}
The rest of this paper is organized as follows: In Sec.\ref{sec:preliminaries}, we explain the preliminaries of database indexes and phantom avoidance techniques. 
In Sec.\ref{sec:our_method}, we introduce our proposed method, \methodname{}.
In Sec.\ref{sec:implementation}, we show our implementation of our method used in the evaluation. In Sec.\ref{sec:evaluation}, we present experimental results on the YCSB benchmark.
In Sec.\ref{sec:relatedwork}, we discuss related work.
In Sec.\ref{sec:conclusion}, we summarize our findings and suggest future work.

\section{Preliminaries}
\label{sec:preliminaries}

\subsection{Index Operations: Point and Range}
\label{sec:interface}

We assume a database system that supports serializable transactions.
A transaction invokes at least one index operation.
An operation may abort in the event of conflicts with other operations. In such cases, the transaction that initiated the operation will also abort.
Tab.~\ref{tab:list_of_apis} shows the minimum operation set that indexes in databases need to process~\cite{10.5555/645913.671312,DBLP:journals/pvldb/ArulrajLML18}.
These operations, excluding \methodname{} operations for now, can be broadly divided into two types: point operations (\Search{}, \Insert{}, and \Delete{}) and range operations (\Scan{}).

A point operation takes a single key, while a \Scan{} takes a search predicate that represents a key range.
A database concurrently processes a mixture of operations in its workload and sends them to its indexes.
For example, in applications such as self-driving vehicles, a database sends a mixture of \Search{}s and \Scan{}s to its indexes;
a \Search{} returns a data item representing a vehicle whose location information is to be updated, and a \Scan{} returns the locations of vehicles nearby.
In another example system that saves server access logs, \Insert{}s are executed at a high rate and \Search{}s and \Scan{}s rarely happen.
Indexes are required to efficiently handle such diverse workloads.

The implementation and performance characteristics of these interfaces vary greatly depending on the underlying data structures.
Hash tables are highly efficient for point operations with an amortized computational complexity of \Oconst{}.
However, they require $O(N)$ full scans for range queries.
B+-trees have a relatively low computational complexity of \OlogN{} for any operation, making them efficient enough for range operations. However, since point operations such as \Search{} are more common in recent applications, the performance penalty of not being \Oconst{} is significant~\cite{DBLP:conf/eurosys/WuNJ19}.

\begin{table*}[t]
  \centering
  \caption{Index operations invoked by the database. \methodname{} operations will be explained in Sec.\ref{subsec:sync_manager}.}
  \begin{tabular}{lll}
    \toprule
    \multicolumn{3}{l}{\small \textsc{Point Operations}}                 \\
      & \Search{}($k$) & Returns the data corresponding to a given key $k$\\
      & \Insert{}($k$) & Creates data corresponding to a given key $k$      \\
      & \Delete{}($k$) & Removes the data corresponding to a given key $k$ \\
      \\
    \multicolumn{3}{l}{\small \textsc{Range Operation}}                   \\
      & \Scan{}($r$)   & Returns all data in the given key range $r$, where $r$ is a single key range (e.g., \texttt{"a" - "z"} )  \\
      \\
    \multicolumn{3}{l}{\small \textsc{\methodname{} Operations}}                   \\
    & \Commit{}() & Notifies the index that a transaction has been committed \\
    & \Abort{}() & Notifies the index that a transaction has been aborted \\
    \bottomrule
  \end{tabular}
  \label{tab:list_of_apis}
\end{table*}

\subsection{Correctness: Linearizability and Serializability}
\label{sec:correctness}

A general design pattern in transactional databases is that indexes are tasked with satisfying two correctness requirements: ensuring linearizability~\cite{DBLP:journals/toplas/HerlihyW90} of operations and eliminating phantoms~\cite{DBLP:journals/cacm/EswarranGLT76}. 
Linearizability is a property of operation scheduling.
It requires that the ordering (scheduling) of operations be consistent with the wall-clock time ordering.
This property precludes \textit{stale reads}, which are reading of old states of the index.
For example, without linearizability, a client of a distributed system may see an old status of directories.

In transactional databases, indexes must also provide a protocol that eliminates phantoms, which is necessary for guaranteeing serializability~\cite{DBLP:journals/cacm/EswarranGLT76,DBLP:conf/vldb/Reimer83}.
Phantoms occur when operations that manipulate data existence (\Insert{}s and \Delete{}s) run concurrently with predicate-based reads (\Scan{}s) and cause \Scan{}s to get inconsistent results~\cite{DBLP:conf/icde/AdyaLO00}.
For example, consider two vehicles running in the same direction in each neighboring lane. One of them executes two \Scan{}s to obtain vehicle data from two lanes 
in a single transaction,
and the other vehicle changes lanes simultaneously (i.e., executes an \Insert{} to the destination lane and \Delete{} from the old lane). 
A phantom causes the vehicle that has performed the scans to read an intermediate state, i.e., it fails to detect the other vehicle in either lane, although it in fact is veering into the same lane.

\subsection{Phantom Avoidance Techniques}
\label{sec:phantom-avoidance}

The techniques for phantom avoidance are classified into two categories.

\subsubsection{B+-Tree}
\paragraph{Next-key locking} 
Databases such as MySQL~\cite{mysql} and DB2~\cite{DB2} adopt the next-key locking method~\cite{DBLP:conf/vldb/Mohan90} in their indexes for phantom avoidance.
This method acquires two locks for each record being accessed. One of them is a lock on the record itself (e.g., a leaf node in the B+-tree) and the second lock is acquired on the pointer to the next record.
\Scan{}s detect the existence of concurrent \Insert{}s and \Delete{}s by finding these locks and wait for them to be unlocked.
Next-key locking has the problem of low concurrency because it acquires two locks for every record it accesses ~\cite{DBLP:conf/sosp/TuZKLM13,DBLP:journals/pvldb/GuoCWQZ19}.

\paragraph{Rescanning} 
Modern transactional database systems employ an optimistic method that avoids locking to fully leverage the parallelism of multiprocessor hardware.
Hekaton~\cite{DBLP:conf/sigmod/DiaconuFILMSVZ13} and Silo~\cite{DBLP:conf/sosp/TuZKLM13} perform a validation process based on the optimistic concurrency control~\cite{DBLP:journals/tods/KungR81}.
This technique performs \Scan{}s without extra locking in the B+-tree and instead stores the ranges of \Scan{}s in transaction-local variables.
It then uses \textit{rescanning} validation where it executes a scan on the same ranges again at commit time.
The validation process checks if the results of the two scans match (i.e., there have been no \Insert{}s and \Delete{}s conflicting with the scan ranges).
Since rescanning does not use locks, its concurrency is higher than next-key locking.
However, the \textit{two-fold} scanning is costly when the read set is large~\cite{DBLP:journals/pvldb/GuoCWQZ19}.

Next-key locking and rescanning are widely adopted by databases with tree indexes because they support both point search and range search, though its point search cost ($O(log N)$) is less efficient than hash search ($O(1)$).

\subsubsection{Precision Locking}
\label{subsubsec:precision-locking}
The second technique uses a special data structure called precision locking~\cite{DBLP:conf/sigmod/JordanBB81}.
Databases such as Hyper~\cite{DBLP:conf/sigmod/0001MK15} and AOCC~\cite{DBLP:journals/pvldb/GuoCWQZ19} employ precision locking~\cite{DBLP:conf/sigmod/JordanBB81}, which is an extension of predicate locking~\cite{DBLP:journals/cacm/EswarranGLT76}.

In precision locking, \Insert{}s, \Delete{}s, \Scan{}s, i.e., operations that can cause phantoms, are guarded with a logical lock associated with the key (or key range) they access.
This lock prevents other transactions from executing a data operation that accesses a conflicting key (or key range.)
This lock is acquired before the actual insert/delete/scan, and released after the transaction that issued the operation terminates.



\begin{table}[tb]
    \centering
    \caption{The \Lu{} and \Lp{} sets in precision locking.}
    \begin{tabular}{c|c}
        \toprule
        \textbf{\Lu{}} & Set of \textbf{updated} keys (\Insert{}s and \Delete{}s) \\
        \textbf{\Lp{}} & Set of read \textbf{predicates} (\Scan{}s) \\
        \bottomrule
    \end{tabular}
    \label{tab:lu-lp}
\end{table}

The logical locks are managed in two sets, {\boldmath \Lu{}} and {\boldmath{}\Lp{}} (terminology in accordance with~\cite{DBLP:journals/cacm/EswarranGLT76}).
These sets have entries that each represent a logical lock.
As shown in Tab. \ref{tab:lu-lp}, {\boldmath \Lu{}} is a set of entries for \Insert{}s and \Delete{}s, and {\boldmath \Lp{}} for \Scan{}s. We (and~\cite{DBLP:journals/cacm/EswarranGLT76}) assume each \Scan{} has a read predicate that is a single key range.
An \Insert{} or \Delete{} first adds its entry that contains the accessed key to \Lu{}, and searches \Lp{} to detect any range that overlaps with the inserted or deleted key. The operation is able to proceed if no conflicts are found, or waits (or gets aborted) otherwise. A \Scan{} follows a similar procedure: it first adds its entry that contains the scanned range (predicate) to \Lp{}, searches \Lu{} for conflicting keys.


\subsection{Properties of Data Structures}
\begin{table*}[tb]
    \centering
    \caption{Summary of data structure properties. $N$ represents the total number of elements stored, $M$ the number of elements returned by a \Scan{}, $|L_{u}|,|L_{p}|$ the sizes of precision locking sets $L_{u},L_{p}$, respectively. Note that range operations (\Scan{}s) in B+-tree entail double the execution costs of a vertical ($O(logN)$) and horizontal ($O(M$)) traversal, because it involves rescanning at commit time for phantom avoidance.}
    \begin{tabular}{c|c|ccc|c}
        \toprule
                 & Operation & Hash & B+-Tree & Precision Locking & \methodname{} \\
                  \midrule
\multirow{2}{*}{Point Operation Costs} & \Search{} & \multirow{2}{*}{\Oconst{}} & \multirow{2}{*}{\OlogN{}} & \multirow{2}{*}{None} & \Oconst{} \\ 
 & \Insert{}/\Delete{} &  &  &  & $O(1) + O(|L_{p}|)$ \\ 
\midrule
Range Operation Costs & \Scan{} & $O(N)$ & $2 \times (O(\log N) + O(M))$ & None & $ O(|L_{u}|) + O(\log N) + O(M)$\\
\midrule
Phantom Avoidance &- & No & Yes & Yes & Yes \\
        \bottomrule
    \end{tabular}
        \label{tab:Summary of Phantom Avoidance}
\end{table*}
In Tab. \ref{tab:Summary of Phantom Avoidance}, we summarize the properties of hash, B+-tree (that avoids phantoms with rescanning), and precision locking from the perspectives of point operation costs, range operation costs, and phantom avoidance.
Hash provides the fastest point operations, while it requires a full search in a \Scan{} and does not provide phantom avoidance.
B+-tree with rescanning provides \OlogN{} point operations, which we have discussed are not optimal (Sec. \ref{sec:introduction}, Issue 1). It provides range operations, but they entail rescanning overhead for phantom avoidance.
Precision locking provides neither point search nor range search, because its data structures (\Lu{} and \Lp{}) themselves do not provide access paths as hash and tree indexing methods do.
%

If we could use hash for point search, tree for range search, and precision locking for phantom avoidance without rescanning, it might be a most efficient structure.

\section{\methodname{}}
\label{sec:our_method}

\begin{figure*}[t]
  \centering
  \includegraphics[width=0.75\textwidth]{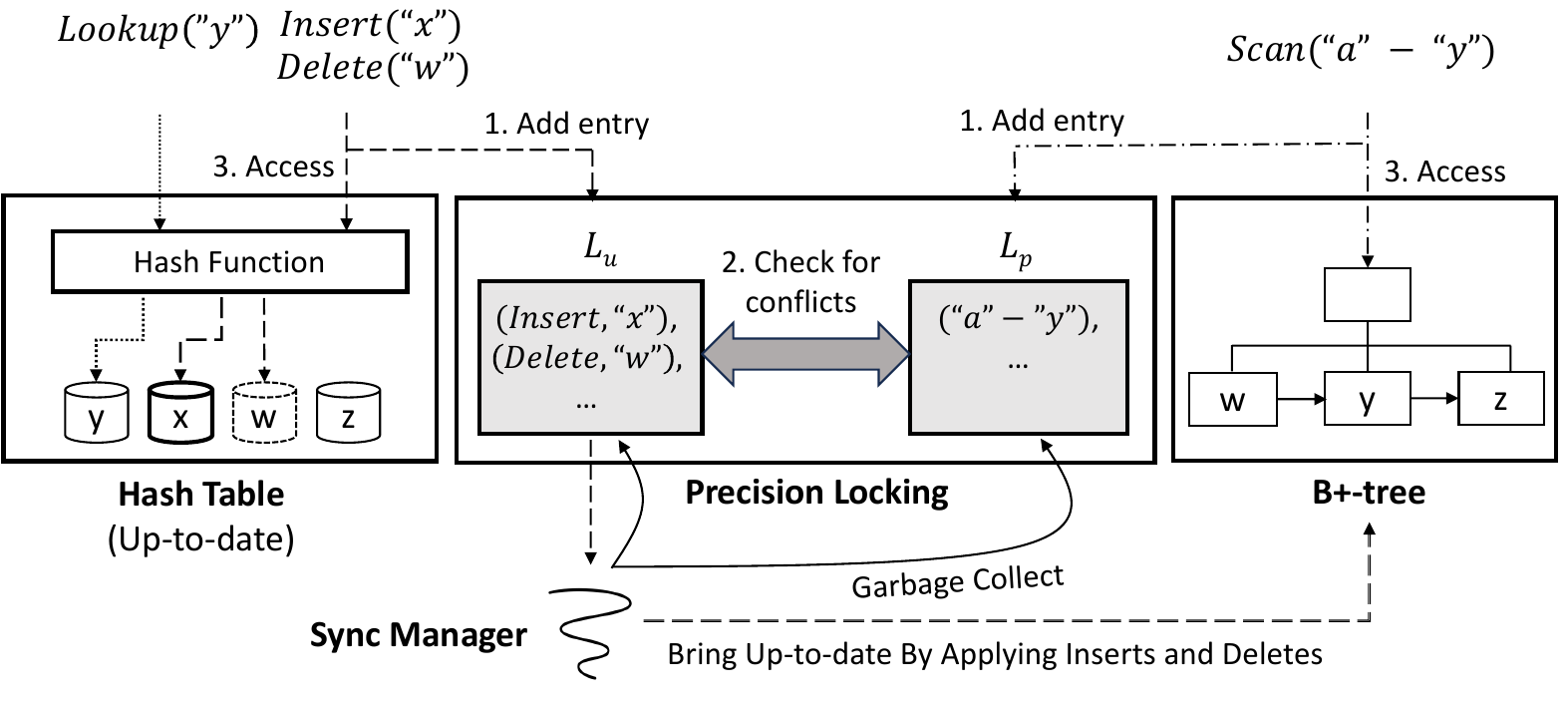}
  \caption{Data structure and control flow of \methodname{}. \methodname{} has an always up-to-date hash table and an asynchronously updated B+-tree. For phantom avoidance, \methodname{} validates \Insert{}ed and \Delete{}d keys and \Scan{}ned key ranges, according to precision locking~\cite{DBLP:journals/cacm/EswarranGLT76}.
  }
  \label{fig:flow}
\end{figure*}

This section proposes a fast index architecture called \methodnameFull{}.
\methodname{} consists of two index structures: a hash table for point operations and a B+tree for \Scan{}s connected via precision locking.
The hash table and B+-tree are integrated with precision locking to eliminate phantoms at a low performance cost (last column of Table \ref{tab:Summary of Phantom Avoidance}).
\methodname{} offers high performance for diverse workloads and guarantees phantom avoidance.

\subsection{Overview}

The notable features of \methodname{}, whose structure is illustrated in Fig.~\ref{fig:flow}, are two-fold.
Firstly, to achieve fast point operations (\Search{}, \Insert{}, \Delete{}), the hash table provides \Oconst{} access paths.
The hash table is paired with a B+-tree, which processes \Scan{}s in \OlogN{}, as in conventional B+-tree indexes.

Secondly, for phantom avoidance, \methodname{} adopts precision locking (Sec. \ref{subsubsec:precision-locking}).
To provide the mutual exclusion in precision locking, operations that can cause phantoms (\Insert{}, \Delete{}, \Scan{}) first go through a conflict check; they search for any conflicting operations being executed. If any conflict is found, the operation and the transaction that issued it is aborted. Only when no conflicts are found do they proceed to access the hash table or the B+-tree.
\Search{}s can directly access the hash table without searching for conflicts because they do not cause phantoms.

\begin{figure}[t]
  \centering
    \includegraphics[height=4cm]{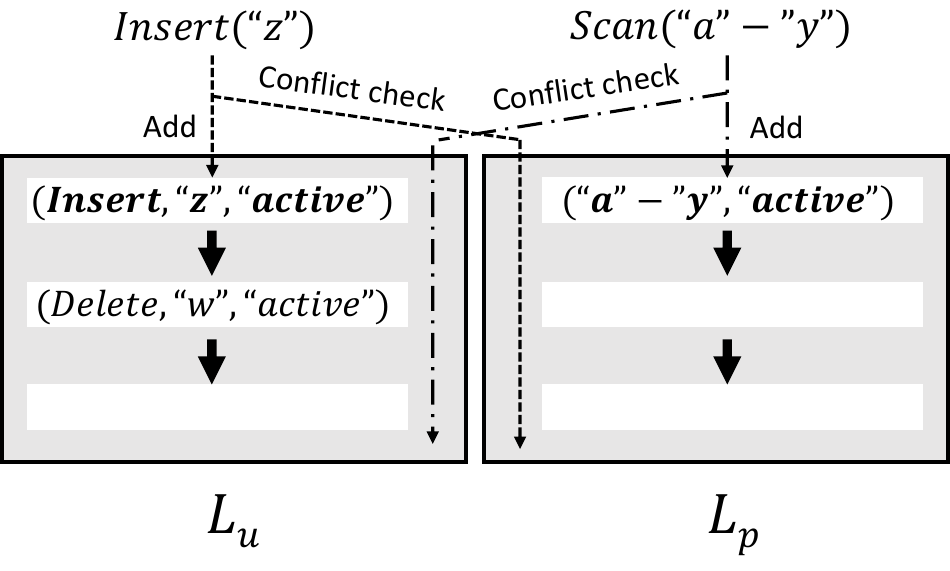}
  \caption{Precision locking design. Each \Lu{} entry has an \Insert{}ed or \Delete{}d key, along with the status of the transaction that issued it (used by \textit{sync manager}, as will be explained in Sec. \ref{subsec:sync_manager}). Each \Lp{} entry has a \Scan{}ned key range and the transaction status.}
  \label{fig:precision_locking}
\end{figure}

The process each operation goes through is as follows. The conflict check in \Insert{}s, \Delete{}s, \Scan{}s are illustrated in Fig. \ref{fig:precision_locking}.

\begin{itemize}
\item{
  \Search{}s only access the hash table, which makes it faster than those on a B+-tree index. This process does not involve access to \Lu{} or \Lp{}.
}
\item{
An \Insert{} or \Delete{} first adds its entry to \Lu{}, and checks for conflicts with \Lp{} entries: if the inserted or deleted key does not overlap with any range stored in \Lp{}, it proceeds to update the hash table. Otherwise, it is aborted.
}
\item {
  A \Scan{} first adds its entry to \Lp{}, and checks for conflicts with \Lu{} entries: if the scan range does not overlap with any key stored in \Lu{}, it proceeds to access the range index. Otherwise, it is aborted. Note that rescanning at commit time is not necessary for phantom avoidance, which makes it faster than a B+-tree index.}
\end{itemize}

Since \Insert{}s and \Delete{}s update the hash table, the hash table is always up-to-date. However, they do not directly update the B+-tree.
To prevent the B+-tree from growing stale,  \methodname{} performs a periodic synchronization where \Insert{}s and \Delete{}s to the hash table are applied to the B+-tree in batches.

\subsection{Guarantee of Correctness}
\label{sec:precision_locking}

\begin{figure*}[t]
  \centering
  \begin{subfigure}[t]{0.48\linewidth}
      \includegraphics[width=0.93\textwidth]{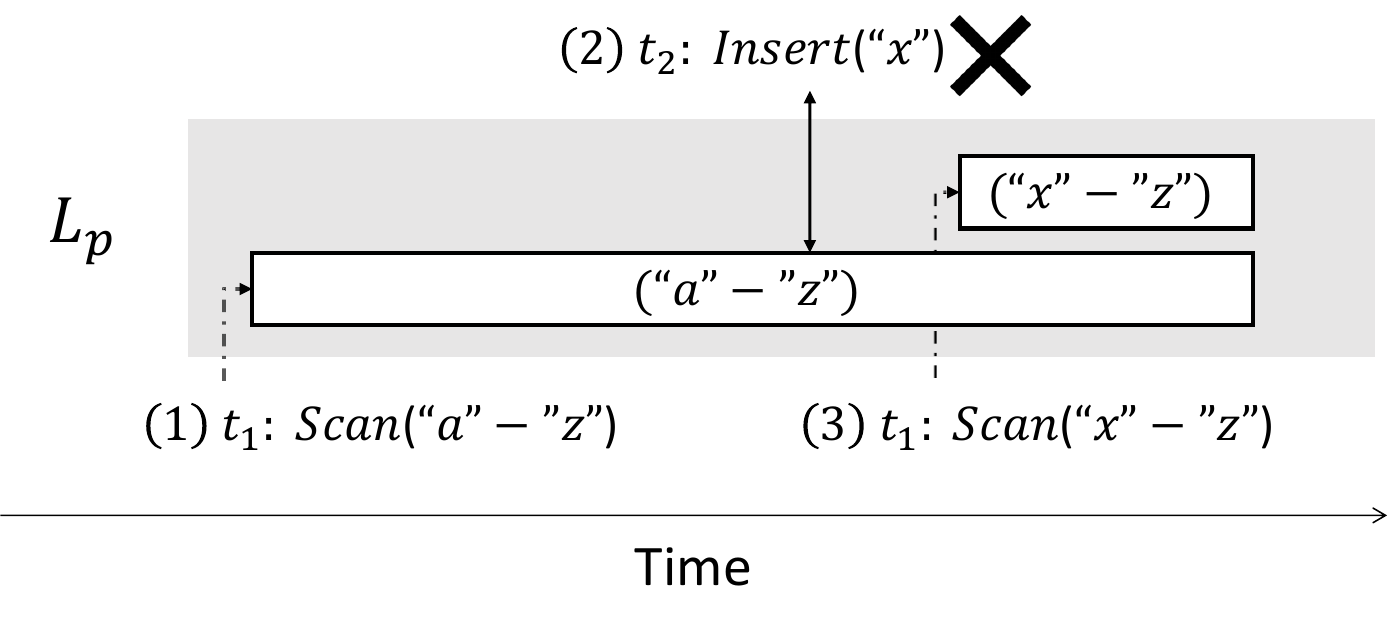}
      \caption{Catching a phantom.}
      \label{fig:phantom_error}
  \end{subfigure}
  \begin{subfigure}[t]{0.48\linewidth}
      \includegraphics[width=0.98\textwidth]{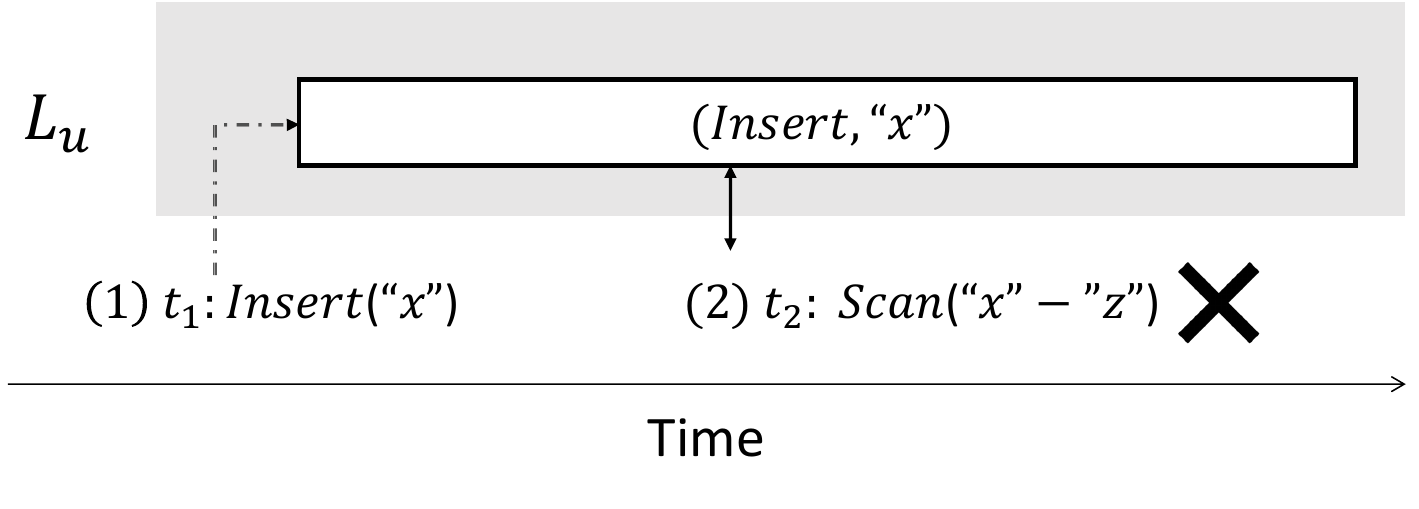}
      \caption{Catching linearizability violation.}
      \label{fig:linearizability_error}
  \end{subfigure}
  \vspace{5pt}
  \caption{Example cases of transaction aborts in \methodname{}. Horizontal axis represents elapsed time. Each box in \Lu{} or \Lp{} denotes the existence of a precision locking entry.}
  \label{fig:aborts}
\end{figure*}

In Sect. \ref{sec:correctness}, we described that the two correctness guarantees that indexes should provide are phantom avoidance and linearizability.
Of these two, phantom avoidance is avoided by forming the mutual exclusion according to precision locking.
\methodname{} leverages precision locking not only for phantom avoidance but also for linearizability.

If \Scan{}s were executed without any restriction, the existence of two index structures could cause stale reads (i.e., linearizability violation).
Since the synchronization of the hash table and B+-tree is periodic, \Scan{}s would miss some recent updates that have been applied only to the hash table but not to the B+-tree.

This situation is prevented by the precision locking scheme. On performing a \Scan{}, whenever there is a preceding \Insert{} or \Delete{} that has not been applied to the B+-tree (a potential stale read situation), \Scan{} will detect its corresponding entry in \Lu{} and abort itself.
The synchronization and lock releasing (garbage collection) algorithm to be explained in Sec. \ref{subsec:sync_manager} ensures that any \Lu{} entry will not be removed until the corresponding \Insert{} or \Delete{} is applied to the B+-tree.

\subsection{Behavior Examples}
Fig.~\ref{fig:aborts} shows two example cases of transaction aborts in \methodname{}.
Both cases show that some \Insert{} and \Scan{} operations are in conflict. Cases (a) and (b) are examples of the detection of a phantom and linearizability violation, respectively.

In case (a), transaction $t_1$ performs two \Scan{}s and transaction $t_2$ performs an \Insert{}.
(1) $t_1$ performs the first \Scan{}, adding an entry to \Lp{}.
(2) $t_2$ tries to execute an \Insert{} to \texttt{x}, but gets aborted because it finds that the entry \texttt{``a'' - ``z''} overlaps with \texttt{x}.
(3) $t_1$ performs the second \Scan{}.
A phantom occurs if the two \Scan{}s and the insert were all successful, because the second \Scan{} (2) would observe the existence of \texttt{x}, while the first \Scan{} of the same transaction (1) would not.

Case (b) depicts a scenario where a conflicting \Scan{} is executed after an \Insert{}.
Here, (1) transaction $t_1$ executes an \Insert{}, and after its completion, (2) transaction $t_2$ executes a  \Scan{} that overlaps with the previous \Insert{}.
Recall that \Insert{}s and \Delete{}s are initially applied only to the hash table. Therefore, if the \Scan{} succeeded, it would not see \texttt{x} \Insert{}ed by $t_1$, even though the \Insert{} had completed at the start of the \Scan{}. This is a stale read (linearizability violation) and is prevented by aborting the \Scan{} in the validation against \Lu{}.

\subsection{Sync Manager}
\label{subsec:sync_manager}

The inconsistency between the hash table and B+-tree is resolved by a dedicated thread, \textit{sync manager}.
This inconsistency resolution task consists of two parts: identifying the difference between the two structures, and bringing the B+-tree up-to-date.

The difference between the two structures is identified by finding all \Insert{}s and \Delete{}s created by committed transactions that have not been applied to the B+-tree.
Since \Insert{} and \Delete{} keys are managed by \Lu{}, each \Lu{} entry is labeled with the execution status of the transaction that created it. 
When an \Lu{} entry is created at the start of an \Insert{} or \Delete{}, it is labeled as \texttt{active}, and then as \texttt{committed} or \texttt{aborted} when the transaction that created it terminates. For the entries to be notified of the transaction termination, we add \Commit{} and \Abort{} to the operation set (Tab. \ref{tab:list_of_apis}), which update the statuses of the entries the transaction has created.

To bring the B+-tree up-to-date, sync manager periodically traverses \Lu{} and finds all \Lu{} entries with a \texttt{committed} status and applies the corresponding inserts and deletes on the B+-tree.
To avoid applying the same insert or delete more than once, an \Lu{} entry is removed once it has been applied to the B+-tree.
Note that \Lp{} is not involved, because it only holds information of read-only operations.

\subsection{Garbage Collection}
Sync manager has another task of garbage collection, i.e., removing precision locking entries ("releasing" of the locks) that no longer potentially cause phantoms or stale reads.
Garbage collection is necessary for optimal performance because retaining \Lu{} and \Lp{} entries long beyond necessity can cause unnecessary aborts, or cause these lists to grow, potentially slowing down conflict detection.
Two correctness requirements govern the deletion of precision locking entries:

\begin{enumerate}
\item {
    For phantom avoidance, an \Lu{} or \Lp{} entry must be deleted only if the transaction that created the entry has terminated (committed or aborted).
}
\item {
    For linearizability, an \Lu{} entry created by a committed transaction must be deleted only if the \Insert{} or \Delete{} operation that created it has been applied to the B+-tree.
}
\end{enumerate}

To meet 1), \methodname{} also labels \Lp{} entries with transaction execution statuses as it does for \Lu{}. Sync manager periodically traverses \Lp{}, as well as \Lu{}, and removes entries with a \texttt{committed} or \texttt{aborted} status.
To meet 2), for \Lu{}, garbage collection needs to work in concert with synchronization: before deleting an \Lu{} entry with a \texttt{committed} status, sync manager must complete applying it to the B+-tree, 

\subsection{Sync Manager Algorithm}
Alg.~\ref{alg:sync_manager} shows an algorithm of sync manager that performs the synchronization of the hash table and the range index while garbage collecting entries. Sync manager periodically inspects \Lu{} and \Lp{} to find entries issued by terminated (with a status \texttt{committed} or \texttt{aborted}) transactions. If an \Lu{} entry is \texttt{committed}, sync manager applies the insert or delete to the B+-tree and then deletes the entry.
\Insert{}s and \Delete{}s on the hash table issued by \texttt{aborted} transactions are rolled back, because they are prematurely made changes.

\begin{algorithm}[t]
      \DontPrintSemicolon
      \SetAlgoLined

      \SetKwInOut{Init}{initialize}
      \SetKwProg{Fn}{Function}{:}{}

      \SetKwFunction{FMain}{Sync\_Manager{}}


      \Fn{\FMain{}}{
            \While{}{
                  
                  \ForEach{
                    entry in $L_u$
                  }{
                    \If{entry has status ``committed''}{
                        B+tree.apply(entry);\;
                        delete entry;\;
                    }
                    \If{entry has status ``aborted''}{
                       hash\_table.rollback(entry);\tcp{\footnotemark[3]}
                       delete entry;\;
                    }
                  }
                  \ForEach{
                    entry in $L_p$
                  }{
                    \If{entry has status ``committed'' or ``aborted''}{
                        delete entry;\;
                    }
                  }
            }
      }
      \caption{Sync manager algorithm.}
      \label{alg:sync_manager}
\end{algorithm}

\subsection{\methodname{} Algorithm}
\begin{algorithm}[t]
  \DontPrintSemicolon
  \SetAlgoLined

  \SetKwInOut{Init}{initialize}
  \SetKwProg{Fn}{Function}{:}{}
  \SetKwFunction{FSearch}{Lookup}
  \SetKwFunction{FInsert}{InsertOrDelete}
  \SetKwFunction{FDelete}{Delete}
  \SetKwFunction{FScan}{Scan}
  \SetKwFunction{FCommit}{Commit}
  \SetKwFunction{FAbort}{Abort}

  \Fn{\FSearch{key}}{
        \KwRet hash\_table.get(key)\;
  }
  \BlankLine
  \Fn{\FInsert{key, value, type: \{``insert'' or ``delete''\}}}{
        entry = $L_u$.add(key, type, ``active'');\;
        entries.add(entry);\;
        \If{key is in some range in $L_p$}{
            \Return{abort}\;
        }
        \If{type is ``insert''}{
            hash\_table.insert(key, value);\;
        }\Else{
            hash\_table.delete(key);\;
        }
  }
  \BlankLine
  \Fn{\FScan{begin, end}}{
        entry = $L_p$.add(begin, end, ``active'');\;
        entries.add(entry);\;
        \If{some key in $L_u$ is in begin-end}{
            \Return{abort}\;
        }
        B+tree.scan(begin, end);\;
  }
  \BlankLine
  \Fn{\FCommit{}}{
    \ForEach{entry in entries}{
        entry.mark(``commit'');
    }
  }
  \BlankLine
  \Fn{\FAbort{}}{
    \ForEach{entry in entries}{
        entry.mark(``aborted'');
    }
  }
  \caption{\methodname{} algorithm. \texttt{entries} is a per-transaction variable that stores the set of \Lu{} and \Lp{} entries the transaction has created.}
  \label{alg:algorithm}
\end{algorithm}

Alg.~\ref{alg:algorithm} shows the algorithm of \methodname{}, which covers all operations shown in Tab.~\ref{tab:list_of_apis}.
\Search{} is the simplest and most efficient operation, accomplished by only accessing the hash table.
An \Insert{} or \Delete{} first adds a lock entry to \Lu{} and then validates against \Lp{} to avoid phantoms.
Once the validation has been successful, it performs the actual changes on the hash table.
A \Scan{} adds a lock entry to \Lp{} and validates against \Lu{} to preclude phantoms and non-linearizable reads. After a successful validation, it accesses the B+-tree.
\methodname{} keeps a per-transaction variable \texttt{entries}, which stores the set of \Lu{} and \Lp{} entries the transaction has created, so that the transaction can identify which entries to mark \texttt{commited} or \texttt{aborted} when \texttt{Commit} or \texttt{Abort} is invoked.
\footnotetext[3]{In "rollback", we delete or insert the entry from the hash table if it is an insert or delete event, respectively.}
\setcounter{footnote}{3}

\section{Implementation}
\label{sec:implementation}

We build a \methodname{} index to compare its performance with a baseline architecture.
As the baseline, we adopt a B+-Tree architecture that avoids phantoms by commit-time rescanning, as done by modern in-memory database systems~\cite{DBLP:journals/pvldb/LarsonBDFPZ11}~\cite{DBLP:conf/sosp/TuZKLM13}.
For both the baseline and \methodname{}, we use Bw-Tree~\cite{DBLP:conf/icde/LevandoskiLS13a}, a modern concurrent B+-tree.

\textbf{B+-tree.}
As a Bw-Tree implementation, we use OpenBwTree, whose source code is available online~\cite{OpenBwTreeAccessed}.
\evalProposal~uses a hash table to process point operations and a Bw-tree to process range operations.
In \evalProposal, Bw-tree does not store data values, only keys, while the hash table stores values or pointers to the values.
Therefore, when a \Scan{} needs to access not only keys but also data values, it uses the obtained keys to execute \Search{} on the hash table.

\textbf{Hash Table.}
\methodname{} has a hash table that optimizes the performance of point operations. We implement a lock-free hash table in C++ that uses open addressing and linear probing~\cite{DBLP:journals/dc/GaoGH05}. It is an array of pointers to data elements, which is updated by atomic CPU instructions. It handles point operations (\Search, \Insert, and \Delete~ operations). 
The amortized calculation cost of this hash table is \Oconst{}, making it efficient for point operations. 

\textbf{Precision Locking.}
We implement \Lu{} and \Lp{} as lock-free singly linked lists~\cite{DBLP:conf/podc/Valois95}, each with a tail pointer.
When an \Insert{}, \Delete{}, or \Scan{} adds an entry (Alg.~\ref{alg:algorithm} Lines 4, 13), it creates a new entry and redirects the tail pointer to it with a compare-and-swap instruction.
When an \Insert{} or \Delete{} searches \Lp{} for conflicts (and when a \Scan{} searches for \Lu{}) (Alg.~\ref{alg:algorithm} Lines 6 and 15), it traverses the entire list, starting from the tail.

\textbf{Sync Manager.} The sync manager (Alg.~\ref{alg:sync_manager}) periodically performs replication of the latest entries from the hash table to the B+-tree. It first traverses \Lu{} from the tail, and if it finds an entry with "committed," it does replication. Entries with "committed" or "aborted" are eliminated by the garbage collection module. Subsequently, it traverses \Lp{} and does only garbage collection since entries in \Lp{} are only \Scan{}, and they do not need to be replicated.

Since the sync manager algorithm does not depend on either the data structures of the hash table or B+-tree, not a single line of code was modified for Bw-tree.

\section{Evaluation}
\label{sec:evaluation}

\subsection{Setup}

\textbf{Workloads.}
We design workloads based on YCSB benchmark\cite{DBLP:conf/cloud/CooperSTRS10}.
We populate the index with 100K keys, each of which is a randomly generated 5-byte string.
We run the following workloads:

\begin{itemize}
    \item \textit{YCSB-A}: This workload contains 100\% \Search{}s. The original YCSB-A consists of 50\% record reads and 50\% record in-place updates. In this experiment, however, the index is oblivious to the processing on record content and makes no distinction between them.

    \item \textit{YCSB-E}: This workload contains 95\%~\Scan{}s and 5\% \Insert{}s. We show a variant of this workload with varying the length of \Scan{}.
  
    \item \textit{Scan-only}: This workload contains 100\% \Scan{}s.
   
    \item \textit{Insert-only}: This workload contains 100\% \Insert{}s.
\end{itemize}

For \Insert{}s and \Search{}s, keys are selected with a uniform distribution.
For \Scan{}s, the start key is selected with a uniform distribution, and the size of the key range is randomly selected (up to 25\% of the entire key space). Each \Scan{} stops after reading 100 keys\footnote{In \evalBaseline{}, the rescan at commit time is done by setting the end key of the read predicate as the last key read in the first scan.}.
Note that YCSB-E is the only workload where phantom anomaly can occur, since it is the only mixture of \Insert{}s and \Scan{}s.

\textbf{Running Environment.}
We evaluate the performance of \evalBaseline{} and \evalProposal{} by building and running them on a server machine with the configuration shown in Tab. \ref{tab:configuration}.
We build \evalProposal{}, the rescanning logic of \evalBaseline{}, and the benchmark logic in C++.
We implement the benchmarks based on the codebase of LineairDB~\cite{lineairdb}, an open-source transactional key-value store.
Index operations and the benchmark logic are executed in a single process.
We instantiate an index for both \evalBaseline{} and \evalProposal{}, allowing multiple threads to access it.
Each thread executes operations (e.g., \Scan{}, \Search{}) as determined by the workloads.
We do not build concurrency control mechanisms on top of the index, i.e., we only measure its performance where only the phantom anomaly is avoided.
We simulate short, single-operation transactions; each operation is followed by a \Commit{} if the operation succeeds or an \Abort{} if it does not.
Each data point in the graphs is derived from an average of three separate runs, with error bars representing the standard deviation.

\begin{table}[tb]
    \centering
    \caption{Running Environment.}
    \begin{tabular}{c|c}
        \toprule
        \textbf{CPU} & Intel Xeon E7-8870 v3 CPU * 4 (72 cores) \\
        \textbf{L3 Cache} & 45MB for each socket \\
        \textbf{DRAM} & 1.5TB \\
        \textbf{OS} & Ubuntu 20.04 \\
        \textbf{Compiler} & Clang 10.0.0 (built with -O3 option) \\
        \bottomrule
    \end{tabular}
    \label{tab:configuration}
\end{table}

\subsection{YCSB-A Result}
\label{sec:ycsb-a}
Fig. \ref{fig:ycsb-a} shows the throughput (committed operations per second) versus the number of threads.
Our method (\evalProposal) shows high scalability and is \YCSBATPS{}x faster than the baseline (\evalBaseline) at \YCSBAThreads{} threads. The improved performance can be attributed to the use of the hash table for handling point operations (\Search), resulting in an amortized \Oconst{} complexity, as opposed to the \OlogN{} complexity of the baseline method, which involves a tree traversal. Please note that Griffin does not access either precision locking or B+tree during \Search{}. Only the hash table completes the processing.

\ExperimentPDF{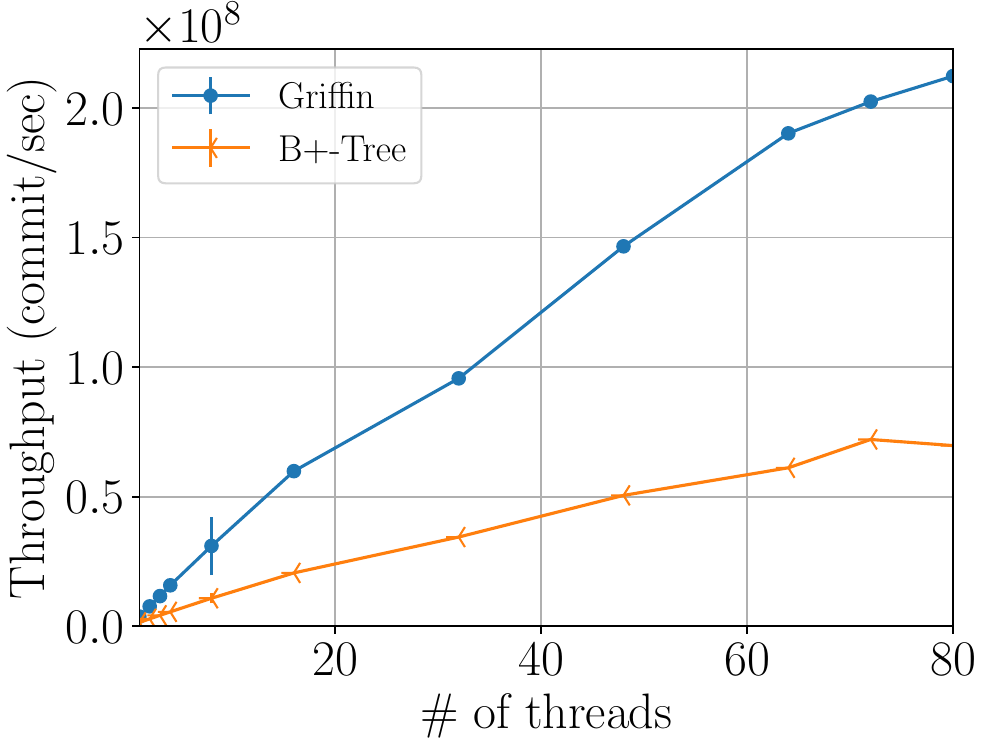}{YCSB-A throughput.}{ycsb-a}
\ExperimentPDF{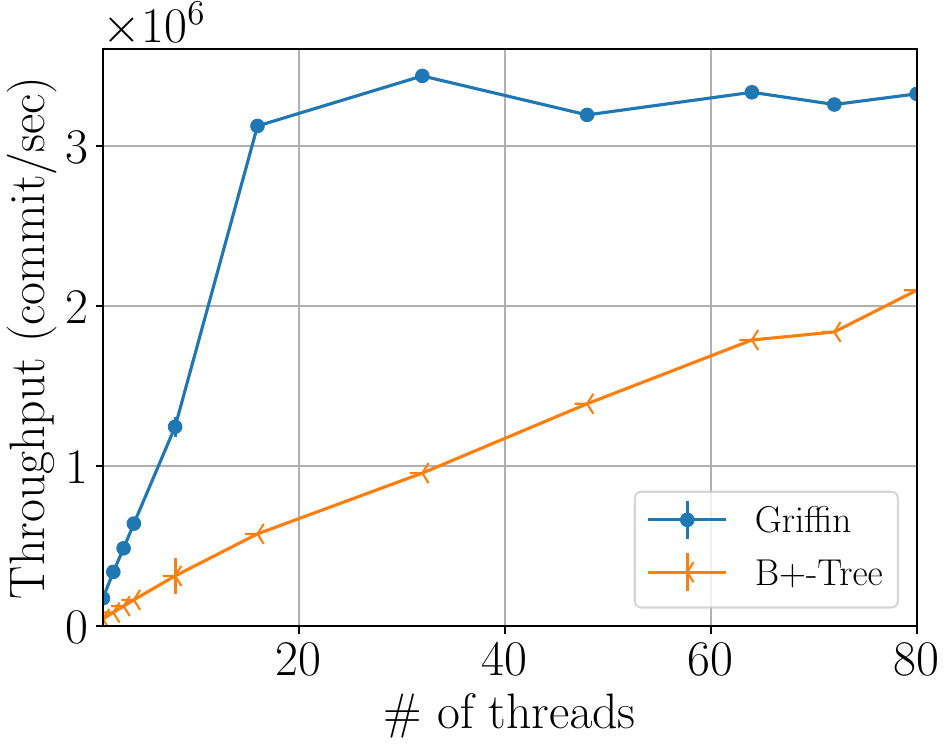}{YCSB-E throughput.}{ycsb-e}
\ExperimentThreePDF{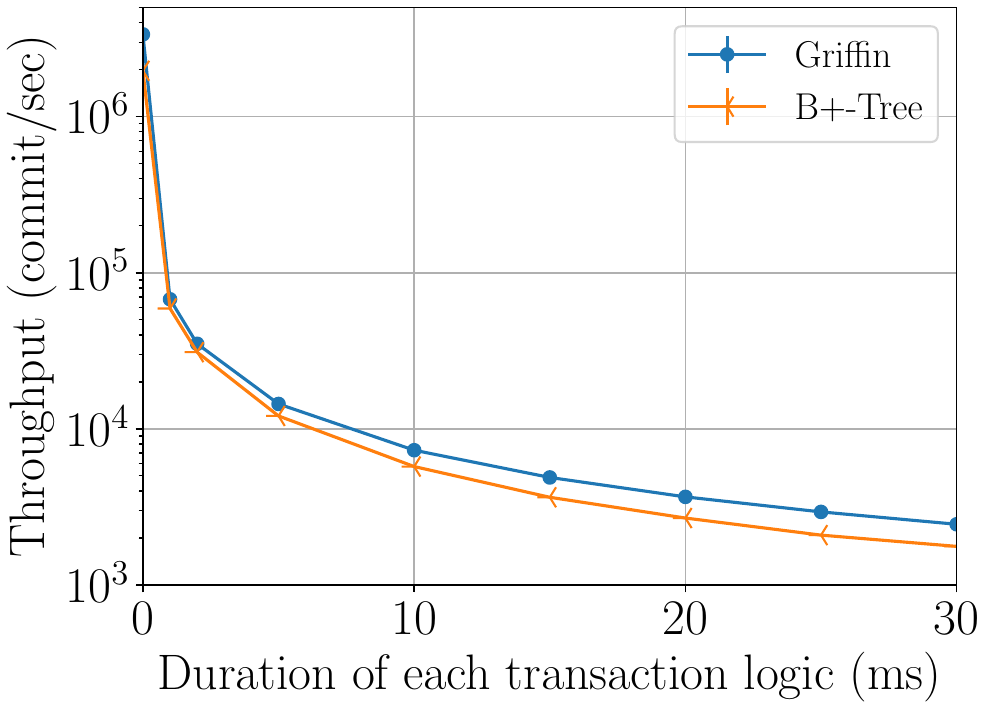}{YCSB-E throughput under variable pause durations. Y-axis is represented on a logarithmic scale.}{ycsb-e-logictime}{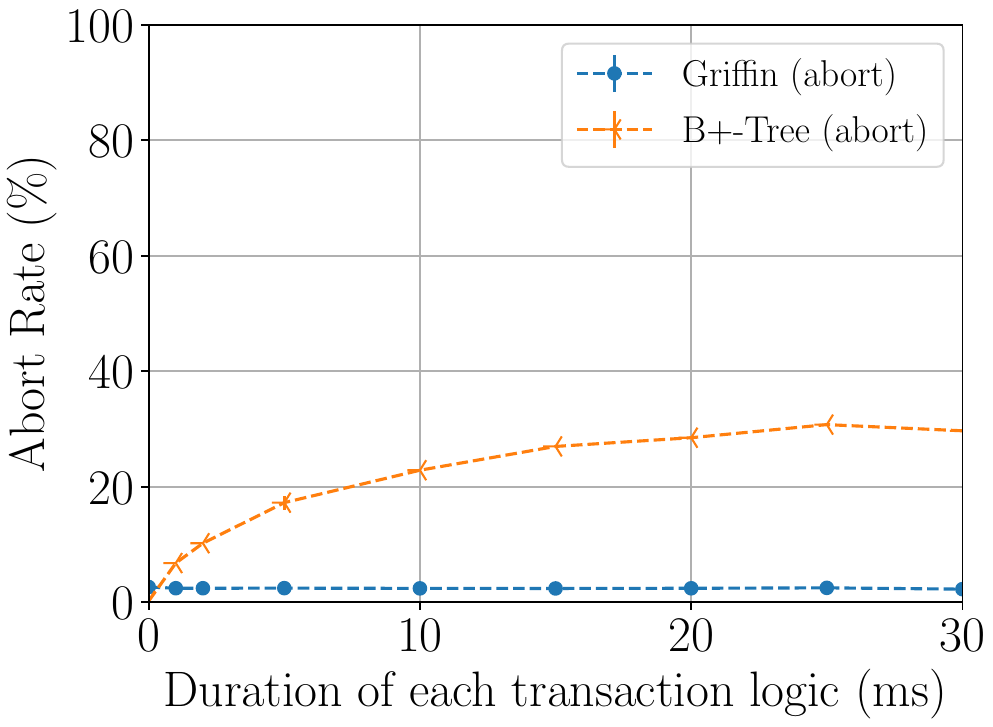}{Abort rates for Fig.~\ref{fig:ycsb-e-logictime}.}{ycsb-e-logictime-aborts}{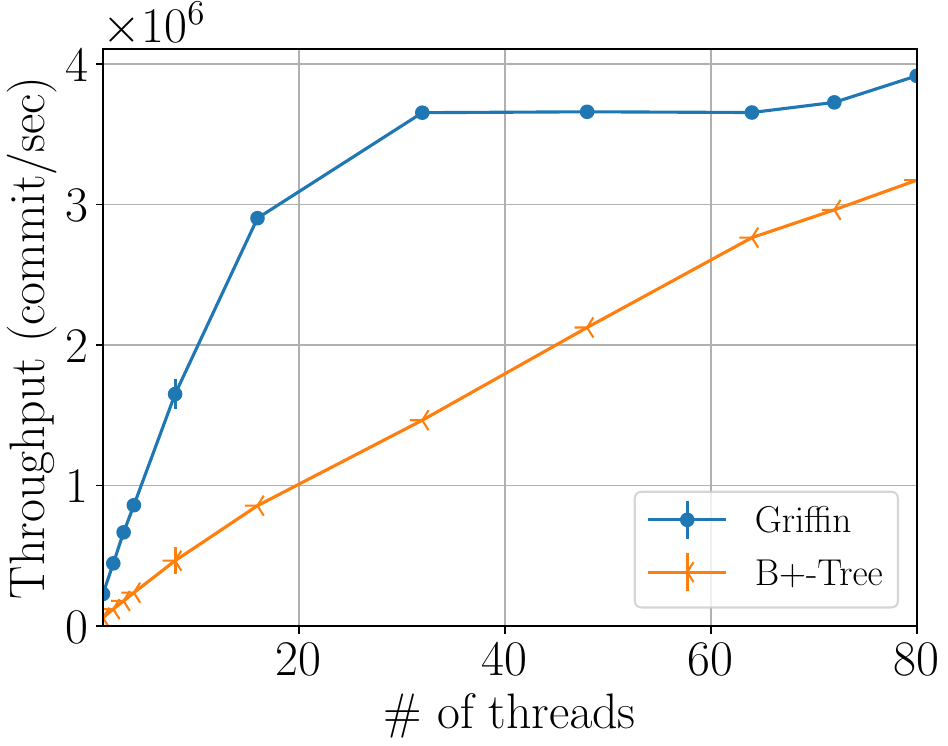}{Scan-only throughput.}{ycsb-scan-only}

\subsection{YCSB-E Result}

\label{sec:ycsb-e}

\textbf{Original Setting.} Fig. \ref{fig:ycsb-e} shows that \evalProposal~demonstrates a \YCSBETPS{}x higher throughput compared to \evalBaseline~at 16 threads and 1.6x higher throughput at 80 threads. These performance gains can be attributed to two factors. Firstly, YCSB-E includes \Insert~operations, which can cause aborts due to phantoms, and \evalProposal~has a lower cost for aborts thanks to precision locking.
\evalBaseline{} always requires twice tree traversals, and it can detect aborts only at the commit time.
On the other hand, \methodname{} traverses the tree at most once, and it can detect aborts before the traversal, which is earlier than that of \evalBaseline{}.
Secondly, \methodname{} has better cache efficiency as it updates the tree structure in a batch every time the sync manager runs, whereas, in the baseline, the tree is updated in each operation.

\textbf{Long Scans.} In this workload, the abort rate for the baseline was close to 0\%. This is because the second validation scan is performed immediately following the first scan. As a result, there are minimal conflicting \Insert{}s executed in between. However, real-world workloads have long scanning transactions~\cite{ByteHTAP, Oze}, whose scan may be followed by additional scans or even joins, resulting in a large gap between the initial scan and the commit-time validation scan.

To consider such cases, we conduct a workload that simulates long transactions. The execution threads are paused for a set duration with a sleep function after each \Scan. After the pause, \evalBaseline~performs a second scan for validation, while \evalProposal~immediately commits. The number of worker threads is fixed at 72. Fig.~\ref{fig:ycsb-e-logictime} displays the performance as the pause duration varies from 1 to 30 milliseconds. As the pause duration increases, the performance of \evalBaseline~deteriorates due to an increase in the number of conflicting inserts and resultant aborts. Fig.~\ref{fig:ycsb-e-logictime-aborts} shows that the abort rate of the experiments in Fig.~\ref{fig:ycsb-e-logictime}. \evalBaseline{} showed an inoperable abort rate of 30\% with longer pause durations. On the other hand, \evalProposal~is more stable.  This is because a \Scan{} in \methodname{} adds an lock entry, preventing itself from being aborted regardless of the pause duration.

\ExperimentPDF{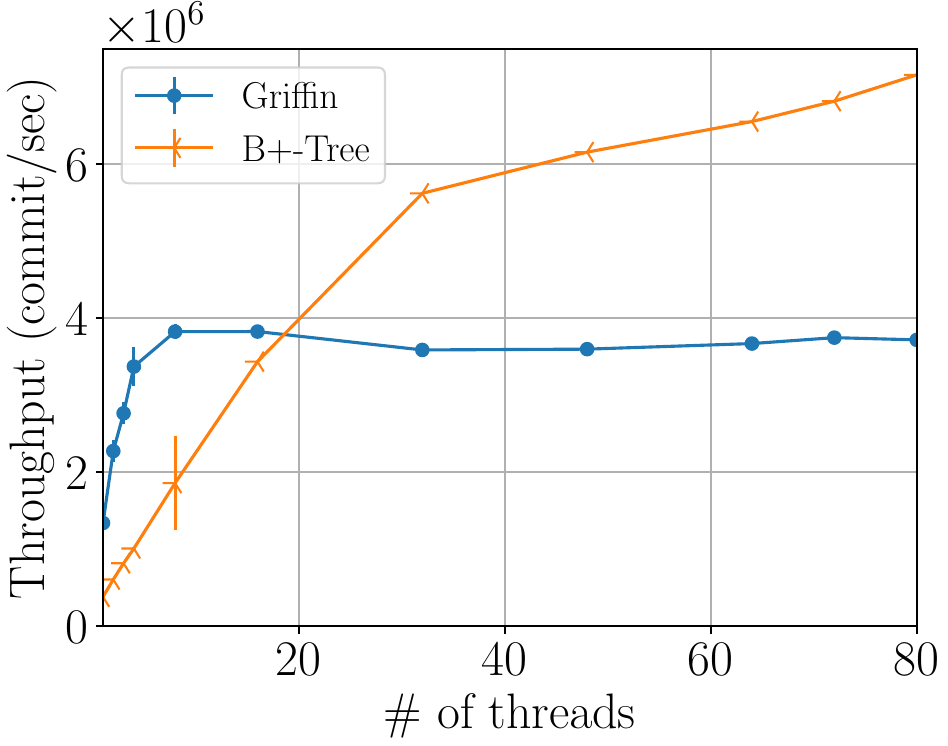}{Insert-only throughput.}{ycsb-insert-only}



\subsection{Scan-only and Insert-only Result}

\textbf{Scan-only.} Fig. \ref{fig:ycsb-scan-only} shows the results of the scan-only workload.
Despite having the same \OlogN{} complexity as \evalBaseline{} for a \Scan, \methodname{} exhibits better performance, which could be attributed to its reduced validation costs (i.e., no rescanning) and improved cache efficiency, which also explains its superior performance in YCSB-E.
\evalProposal{}'s scan for this workload consists of adding an entry to \Lp{}, validating an empty \Lu{}, and a single scan.
In contrast, \evalBaseline{} performs two scans for rescanning.
In general, pushback to the lock-free list (one CAS in the best case) is significantly faster than scan ($N+1$ memory accesses and compare, where $N$ is the number of entries in the range). This difference is more notable in cases where disk accesses occur after the index retrieval.
However, the performance gains are not as substantial as in YCSB-E as there are no aborts.

\textbf{Insert-only.}
This workload represents a scenario where many new records are inserted without being read or updated.
The cost of \Insert{} to the B+-tree is \OlogN{} for both \evalBaseline{}. 
Griffin's overhead is $O(1)$ for the hash table and for the addition to \Lu{}. 
Since the distribution is uniform (the skew parameter is 0), almost no contention occurs for adding a new entry to the tree, and thus it is efficient.
On the other hand, adding entries to \Lu{} severely limits efficiency because all the concurrent
\Insert{}s collide with the tail pointer of the same linked list.
Thus, \evalBaseline{} makes inserts directly to the tree without any collision.
The results of Insert-only are shown in Fig.~\ref{fig:ycsb-insert-only}.
\evalProposal{} outperforms the baseline until reaching saturation around eight threads. 
However, with larger thread counts, the baseline exhibits up to approximately 1.9x better performance at 80 threads.
At the higher thread counts where \evalProposal{} underperforms \evalBaseline{}, the primary bottleneck is identified as the entry addition process of \Lu{}, which involves updates of atomic CPU variables in a lock-free linked list.

It is known that centralization can be a bottleneck in transaction processing~\cite{DBLP:journals/pvldb/TanabeHKT20}. 
By decentralizing the \Lu{} currently implemented in the single lock-free list, or using a batched scheme~\cite{DBLP:conf/sigmod/ThomsonDWRSA12,DBLP:journals/ijnc/UchidaK24}, \methodname{} would alleviate the contentions and may outperform the baseline even in this situation. This is left for future work.

\subsection{Summary}
\methodname{} offered a direct access path to the hash table for point operations, and it exhibited about 3.1x better performance than B+-tree in \Search{}s, but had scalability problems in \Insert{}s.
For \Scan{}s and the mixture of \Scan{}s and \Insert{}s, \methodname{} provides an effective phantom avoidance approach, precision locking, which eliminates rescanning in B+-tree and results in about 5.4x better performance.
We observe that the advantages of \methodname{} solve the issues of B+-tree described in Sec.~\ref{sec:introduction} in 1) improving the performance of \Search{}s, 3) increasing the throughput of \Scan{}s, which we believe are desirable properties for a wide range of practical applications.


\section{Related Work}
\label{sec:relatedwork}

\textbf{Phantom avoidance via tree index.} To prevent phantoms, various systems add synchronization mechanisms to their indexes. MySQL~\cite{mysql} and DB2~\cite{DB2} use \textit{next-key locking}, which synchronizes conflicting operations by acquiring locks for both the target and subsequent records. Hekaton~\cite{DBLP:journals/pvldb/LarsonBDFPZ11,DBLP:conf/icde/LevandoskiLS13a} employs \textit{rescanning}, where \Scan~is repeated at commit time. Silo~\cite{DBLP:conf/sosp/TuZKLM13} implements a rescanning method optimized for Masstree\cite{DBLP:conf/eurosys/MaoKM12}.

\textbf{Phantom avoidance via predicate-based approach.} Another approach to phantom avoidance is \textit{predicate locking}~\cite{DBLP:journals/cacm/EswarranGLT76,10.5555/645911.671118}, the prior work most relevant to precision locking. Unlike next-key locking, predicate locking does not rely on synchronization in indexes.
Transactions post predicates as they execute, which serve as \textit{predicate locks} in detecting conflicts between transactions and determining which transactions to block.
    However, this method has not been used in practice because detecting conflicts between general predicates is NP-hard~\cite{}. Precision locking~\cite{DBLP:conf/sigmod/JordanBB81} reduces the computational cost by limiting the operation set to \Insert, \Delete~and \Scan, and predicates to ranges.
\methodname{} uses precision locking as a means of phantom avoidance and ensuring linearizability.

\textbf{Index with multiple data structures.} HydraList \cite{DBLP:journals/pvldb/MathewM20} is similar to \methodname{} in that it divides the index into multiple data structures. HydraList splits the index into a search layer with low update frequency and a data layer with frequent updates, each implemented using a different data structure.
A common aspect of HydraList and \methodname{} is that it synchronizes the search layer and data layer asynchronously to ensure consistency. However, HydraList is not designed for transaction processing and does not propose a way to avoid phantoms.

ScaleDB~\cite{DBLP:conf/osdi/MehdiH0A23} also adopts a strategy akin to \methodname{}: delaying updates to range indexes, buffering them in hash tables (indexlets), and merging periodically.
The indexlets also accept point operations as the hash table in \methodname{} does.
The difference between ScaleDB and \methodname{} lies in the fairness of operations: ScaleDB prioritizes inserts and deletes than scans, possibly leading to unfair aborts of scans in some workloads that include long-running transactions~\cite{Oze,ByteHTAP}.
In ScaleDB, inserts or deletes do not get aborted. On the other hand, in \methodname{}, an insert or delete gives way to concurrent scans that have posted their precision lock entries to \Lp{} before the insert or delete does to \Lu{}. ScaleDB does not have its counterpart of \Lp{} in precision locking.

\section{Conclusions}
\label{sec:conclusion}

We propose \methodname{}, a hybrid index architecture for transactional databases that are designed to improve point operation performance and avoid phantoms with low overhead while providing linearizable operations.
\methodname{} incorporates two data structures: a hash table that provides \Oconst{} point operations and a range index that is typically a B+-tree variant. \methodname{} avoids phantoms with low overhead by employing precision locking.
\methodname{} synchronizes the two structures periodically while eliminating linearizability violations by leveraging precision locking. As a result, \methodname{} offers a transparent interface of a single index with phantom avoidance and linearizable operations, despite its hybrid architecture.
We showed empirically that \methodname{}, compared to the conventional architecture of a B+-tree, results in a marked performance improvement: a \YCSBATPS{}x peak throughput for a point operation dominant workload and \YCSBETPS{}x for a range operation dominant workload.

Despite these promising results, several avenues for future work remain.
Firstly, under a very update-heavy workload, we observed that \Lu{} and \Lp{} could be a major source of contention that could bottleneck the performance. Exploring more scalable implementations of precision locking would be beneficial.
Secondly, while we adopted a B+-tree to handle \Scan{}s, evaluating the performance characteristics of \methodname{} with alternative range index structures would be valuable. 
Since precision locking is oblivious to the index data structure, we can easily use any ordered concurrent index structure, including a skip list~\cite{skiplist}, LSM-tree~\cite{lsm-tree}, etc. 



\bibliographystyle{ieeetr}
\bibliography{bibliography}

\begin{thebibliography}{10}

\bibitem{fastquery}
J.~Chou, K.~Wu, and Prabhat, ``Fastquery: A parallel indexing system for
  scientific data,'' in {\em 2011 IEEE International Conference on Cluster
  Computing}, pp.~455--464, 2011.

\bibitem{insitu}
J.~Kim, H.~Abbasi, L.~Chacón, C.~Docan, S.~Klasky, Q.~Liu, N.~Podhorszki,
  A.~Shoshani, and K.~Wu, ``Parallel in situ indexing for data-intensive
  computing,'' in {\em 2011 IEEE Symposium on Large Data Analysis and
  Visualization}, pp.~65--72, 2011.

\bibitem{hopsfs}
S.~Niazi, M.~Ismail, S.~Haridi, J.~Dowling, S.~Grohsschmiedt, and
  M.~Ronstr\"{o}m, ``Hopsfs: scaling hierarchical file system metadata using
  newsql databases,'' in {\em Proceedings of the 15th Usenix Conference on File
  and Storage Technologies}, pp.~89--103, 2017.

\bibitem{cfs}
Y.~Wang, Y.~Wu, C.~Li, P.~Zheng, B.~Cao, Y.~Sun, F.~Zhou, Y.~Xu, Y.~Wang, and
  G.~Xie, ``Cfs: Scaling metadata service for distributed file system via
  pruned scope of critical sections,'' in {\em Proceedings of the Eighteenth
  European Conference on Computer Systems}, EuroSys '23, p.~331–346, 2023.

\bibitem{DBLP:journals/ngc/TatebeHS10}
O.~Tatebe, K.~Hiraga, and N.~Soda, ``Gfarm grid file system,'' {\em New Gener.
  Comput.}, vol.~28, no.~3, pp.~257--275, 2010.

\bibitem{DBLP:conf/micro/KocberberGPFLR13}
Y.~O. Ko{\c{c}}berber, B.~Grot, J.~Picorel, B.~Falsafi, K.~T. Lim, and
  P.~Ranganathan, ``Meet the walkers: accelerating index traversals for
  in-memory databases,'' in {\em {MICRO}}, pp.~468--479, {ACM}, 2013.

\bibitem{dbdbio}
dbdb.io, ``{D}atabase of {D}atabases.'' \url{https://dbdb.io/browse}.
\newblock [Accessed 19-Sep-2022].

\bibitem{DBLP:books/daglib/0023376}
T.~H. Cormen, C.~E. Leiserson, R.~L. Rivest, and C.~Stein, {\em Introduction to
  Algorithms, 3rd Edition}.
\newblock {MIT} Press, 2009.

\bibitem{DBLP:journals/cacm/EswarranGLT76}
K.~P. Eswaran, J.~Gray, R.~A. Lorie, and I.~L. Traiger, ``The notions of
  consistency and predicate locks in a database system,'' {\em Commun. {ACM}},
  vol.~19, no.~11, pp.~624--633, 1976.

\bibitem{DBLP:journals/pvldb/LarsonBDFPZ11}
P.~Larson, S.~Blanas, C.~Diaconu, C.~Freedman, J.~M. Patel, and M.~Zwilling,
  ``High-performance concurrency control mechanisms for main-memory
  databases,'' {\em Proc. {VLDB} Endow.}, 2011.

\bibitem{DBLP:conf/sosp/TuZKLM13}
S.~Tu, W.~Zheng, E.~Kohler, B.~Liskov, and S.~Madden, ``Speedy transactions in
  multicore in-memory databases,'' in {\em {SOSP}}, pp.~18--32, 2013.

\bibitem{DBLP:conf/sigmod/JordanBB81}
J.~R. Jordan, J.~Banerjee, and R.~B. Batman, ``Precision locks,'' in {\em
  SIGMOD Conf.}, pp.~143--147, 1981.

\bibitem{DBLP:journals/toplas/HerlihyW90}
M.~Herlihy and J.~M. Wing, ``Linearizability: {A} correctness condition for
  concurrent objects,'' {\em {ACM} Trans. Program. Lang. Syst.}, vol.~12,
  no.~3, pp.~463--492, 1990.

\bibitem{DBLP:journals/dc/GaoGH05}
H.~Gao, J.~F. Groote, and W.~H. Hesselink, ``Lock-free dynamic hash tables with
  open addressing,'' {\em Distributed Comput.}, vol.~18, no.~1, pp.~21--42,
  2005.

\bibitem{DBLP:conf/sigmod/WangPLLZKA18}
Z.~Wang, A.~Pavlo, H.~Lim, V.~Leis, H.~Zhang, M.~Kaminsky, and D.~G. Andersen,
  ``Building a bw-tree takes more than just buzz words,'' in {\em {SIGMOD}
  Conf.}, pp.~473--488, 2018.

\bibitem{DBLP:conf/icde/LevandoskiLS13a}
J.~J. Levandoski, D.~B. Lomet, and S.~Sengupta, ``The bw-tree: {A} b-tree for
  new hardware platforms,'' in {\em {ICDE}}, pp.~302--313, 2013.

\bibitem{10.5555/645913.671312}
T.~J. Lehman and M.~J. Carey, ``A study of index structures for main memory
  database management systems,'' in {\em VLDB}, p.~294–303, 1986.

\bibitem{DBLP:journals/pvldb/ArulrajLML18}
J.~Arulraj, J.~J. Levandoski, U.~F. Minhas, and P.~Larson, ``Bztree: {A}
  high-performance latch-free range index for non-volatile memory,'' {\em
  PVLDB}, vol.~11, no.~5, pp.~553--565, 2018.

\bibitem{DBLP:conf/eurosys/WuNJ19}
X.~Wu, F.~Ni, and S.~Jiang, ``Wormhole: {A} fast ordered index for in-memory
  data management,'' in {\em Proceedings of the Fourteenth EuroSys Conference
  2019, Dresden, Germany, March 25-28, 2019}, pp.~18:1--18:16, {ACM}, 2019.

\bibitem{DBLP:conf/vldb/Reimer83}
M.~Reimer, ``Solving the phantom problem by predicative optimistic concurrency
  control,'' in {\em VLDB}, pp.~81--88, 1983.

\bibitem{DBLP:conf/icde/AdyaLO00}
A.~Adya, B.~Liskov, and P.~E. O'Neil, ``Generalized isolation level
  definitions,'' in {\em ICDE}, pp.~67--78, 2000.

\bibitem{mysql}
MySQL, ``{M}y{S}{Q}{L} 8.0 {R}eference {M}anual :: 15.7.4 {P}hantom {R}ows.''
  \url{https://dev.mysql.com/doc/refman/8.0/en/innodb-next-key-locking.html}.
\newblock [Accessed 19-Sep-2022].

\bibitem{DB2}
IBM, ``Db2 version 10.1 for linux, next-key locking.''
  \url{https://www.ibm.com/docs/en/db2/10.1.0?topic=management-next-key-locking}.
\newblock [Accessed 19-Sep-2022].

\bibitem{DBLP:conf/vldb/Mohan90}
C.~Mohan, ``{ARIES/KVL:} {A} key-value locking method for concurrency control
  of multiaction transactions operating on b-tree indexes,'' in {\em VLDB},
  pp.~392--405, 1990.

\bibitem{DBLP:journals/pvldb/GuoCWQZ19}
J.~Guo, P.~Cai, J.~Wang, W.~Qian, and A.~Zhou, ``Adaptive optimistic
  concurrency control for heterogeneous workloads,'' {\em PVLDB}, vol.~12,
  no.~5, pp.~584--596, 2019.

\bibitem{DBLP:conf/sigmod/DiaconuFILMSVZ13}
C.~Diaconu, C.~Freedman, E.~Ismert, P.~Larson, P.~Mittal, R.~Stonecipher,
  N.~Verma, and M.~Zwilling, ``Hekaton: {SQL} server's memory-optimized {OLTP}
  engine,'' in {\em SIGMOD Conf.}, pp.~1243--1254, 2013.

\bibitem{DBLP:journals/tods/KungR81}
H.~T. Kung and J.~T. Robinson, ``On optimistic methods for concurrency
  control,'' {\em {ACM} Trans. Database Syst.}, pp.~213--226, 1981.

\bibitem{DBLP:conf/sigmod/0001MK15}
T.~Neumann, T.~M{\"{u}}hlbauer, and A.~Kemper, ``Fast serializable
  multi-version concurrency control for main-memory database systems,'' in {\em
  {SIGMOD} Conf.}, pp.~677--689, 2015.

\bibitem{OpenBwTreeAccessed}
GitHub, ``wangziqi2013/{B}w{T}ree: {A}n open sourced implementation of
  {B}w-{T}ree in {S}{Q}{L} {S}erver {H}ekaton.''
  \url{https://github.com/wangziqi2013/BwTree}.
\newblock [Accessed 19-Sep-2022].

\bibitem{DBLP:conf/podc/Valois95}
J.~D. Valois, ``Lock-free linked lists using compare-and-swap,'' in {\em PODC},
  pp.~214--222, 1995.

\bibitem{DBLP:conf/cloud/CooperSTRS10}
B.~F. Cooper, A.~Silberstein, E.~Tam, R.~Ramakrishnan, and R.~Sears,
  ``Benchmarking cloud serving systems with {YCSB},'' in {\em SoCC},
  pp.~143--154, 2010.

\bibitem{lineairdb}
GitHub, ``{L}ineair{D}{B}/{L}ineair{D}{B}: {C}++ fast transactional key-value
  storage.'' \url{https://github.com/LineairDB/LineairDB}.
\newblock [Accessed 19-Sep-2022].

\bibitem{ByteHTAP}
J.~Chen, Y.~Ding, Y.~Liu, F.~Li, L.~Zhang, M.~Zhang, K.~Wei, L.~Cao, D.~Zou,
  Y.~Liu, L.~Zhang, R.~Shi, W.~Ding, K.~Wu, S.~Luo, J.~Sun, and Y.~Liang,
  ``Bytehtap: bytedance's htap system with high data freshness and strong data
  consistency,'' {\em Proc. VLDB Endow.}, vol.~15, p.~3411–3424, aug 2022.

\bibitem{Oze}
J.~Nemoto, T.~Kambayashi, T.~Hoshino, and H.~Kawashima, ``Oze: Decentralized
  graph-based concurrency control for real-world long transactions on bom
  benchmark,'' {\em CoRR}, vol.~abs/2210.04179, 2022.

\bibitem{DBLP:journals/pvldb/TanabeHKT20}
T.~Tanabe, T.~Hoshino, H.~Kawashima, and O.~Tatebe, ``An analysis of
  concurrency control protocols for in-memory database with ccbench,'' {\em
  PVLDB}, vol.~13, no.~13, pp.~3531--3544, 2020.

\bibitem{DBLP:conf/sigmod/ThomsonDWRSA12}
A.~Thomson, T.~Diamond, S.~Weng, K.~Ren, P.~Shao, and D.~J. Abadi, ``Calvin:
  fast distributed transactions for partitioned database systems,'' in {\em
  SIGMOD Conf.} (K.~S. Candan, Y.~Chen, R.~T. Snodgrass, L.~Gravano, and
  A.~Fuxman, eds.), pp.~1--12, {ACM}, 2012.

\bibitem{DBLP:journals/ijnc/UchidaK24}
M.~Uchida and H.~Kawashima, ``{CLMD:} making lock manager predictable and
  concurrent for deterministic concurrency control,'' {\em Int. J. Netw.
  Comput.}, vol.~14, no.~1, pp.~81--92, 2024.

\bibitem{DBLP:conf/eurosys/MaoKM12}
Y.~Mao, E.~Kohler, and R.~T. Morris, ``Cache craftiness for fast multicore
  key-value storage,'' in {\em Eurosys}, pp.~183--196, 2012.

\bibitem{10.5555/645911.671118}
M.~Reimer, ``Solving the phantom problem by predicative optimistic concurrency
  control,'' in {\em Proceedings of the 9th International Conference on Very
  Large Data Bases}, VLDB '83, (San Francisco, CA, USA), p.~81–88, Morgan
  Kaufmann Publishers Inc., 1983.

\bibitem{DBLP:journals/pvldb/MathewM20}
A.~Mathew and C.~Min, ``Hydralist: {A} scalable in-memory index using
  asynchronous updates and partial replication,'' {\em PVLDB}, vol.~13, no.~9,
  pp.~1332--1345, 2020.

\bibitem{DBLP:conf/osdi/MehdiH0A23}
S.~A. Mehdi, D.~Hwang, S.~Peter, and L.~Alvisi, ``Scaledb: {A} scalable,
  asynchronous in-memory database,'' in {\em {OSDI}}, pp.~361--376, {USENIX}
  Association, 2023.

\bibitem{skiplist}
M.~Fomitchev and E.~Ruppert, ``Lock-free linked lists and skip lists,'' in {\em
  {PODC} 2004}, pp.~50--59, 2004.

\bibitem{lsm-tree}
P.~E. O'Neil, E.~Y.~C. Cheng, D.~Gawlick, and E.~J. O'Neil, ``The
  log-structured merge-tree (lsm-tree),'' {\em Acta Informatica}, vol.~33,
  pp.~351--385, 1996.

\end{thebibliography}

\end{document}